# Electromagnetic fields and charges in '3+1' spacetime derived from symmetry in '3+3' spacetime

J.E. Carroll

Dept. of Engineering, University of Cambridge, CB2 1PZ

E-mail: jec@eng.cam.ac.uk

**Abstract:**

Maxwell's classic equations with fields, potentials, positive and negative charges in '3+1' spacetime are derived solely from the symmetry that is required by the Geometric Algebra of a '3+3' spacetime. The observed direction of time is augmented by two additional orthogonal temporal dimensions forming a complex plane where *i* is the bi-vector determining this plane. Variations in the 'transverse' time determine a zero rest mass. Real fields and sources arise from averaging over a contour in transverse time. Positive and negative charges are linked with poles and zeros generating traditional plane waves with 'right-handed' **E**, **B**, with a direction of propagation **k**. Conjugation and space-reversal give 'left-handed' plane waves without any net change in the chirality of '3+3' spacetime. Resonant wave-packets of left- and right- handed waves can be formed with eigen frequencies equal to the characteristic frequencies $(N+\tfrac{1}{2})f_O$ that are observed for photons.

# 1. Introduction

There are many and varied starting points for derivations of Maxwell's equations. For example Kobe uses the gauge invariance of the Schrödinger equation [**1**] while Feynman, as reported by Dyson, starts with Newton's classical laws along with quantum non-commutation rules of position and momentum [**2**]. Non-commutivity is also important in Bracken's Poisson bracket formalism [**3**]. Silagadze has extended the Feynman-Dyson derivation to cover extra spatial dimensions [**4**]. Extra dimensions allowing for charge/mass ratio and curvature are also introduced in a paper leading to Maxwell's equations via Kaluza's theory in generalised co-ordinates [**5**]. Gersten's derivation starts from symmetry requirements that are similar to those used to form Dirac's equation [**6**]. This latter approach has been generalised by Dvoeglazov adding additional spatial dimensions that are considered essential to understanding quantised fields [**7**].

The present paper derives Maxwell's equations purely from the symmetry of '3+1' spacetime embedded in a '3+3' spacetime. Electromagnetic force equations are not derived here but, given Maxwell's equations, the reader could refer to Jefimenko [**8**]. The motivation for adding two temporal dimensions arose from Cramer's transactional quantum theory where signals are said to circulate forward and backward in time [**9**]. However in fluid flow, circulation or vortices are a 'peculiar and characteristic glory' of *three-dimensions* [**10**]. These two theories taken together suggested that three temporal dimensions should be re-investigated. A conference paper started this task [**11**], but has a number of faults: potentials have to be assumed; the source currents have zero curl; positive and negative charges are not identified and the left-handed fields, discussed later, are also not recognised. These deficiencies are addressed here.

Three dimensional time has been considered extensively in relation to superluminal particles [**12**, **13**, **14**, **15**, **16**]. However 3-d time raises many valid criticisms [**17**, **18**, **19**, **20**, **21**]. The key parts of these criticisms arise from the assumption that time is homogeneous in the same way as space. As a result, reciprocal time would allow energy to have three 'interchangeable' dimensions. Then just as the magnitude of two sides of a triangle can exceed that of the third, so a particle with one component of energy could split (or combine) into (or from) two particles whose combined energies exceeds that of the initial particle. Strnad reports that the experimental evidence is against such three-dimensional time [**22**].

There are several other 6 dimensional theories but keeping a single time and adding extra *spatial* dimensions, for example to allow for curvature of space or for mass normalisation [**5, 23**]. Even in multi-dimensional string and brane theories there is caution about having additional dimensions of time perhaps because, according to Tegmark, the stability of physical interactions can be compromised [**24**]. Never the less there are suggestions in brane theory that extra time-like dimensions could be considered [**25, 26, 27**]. Discussions on brane and string theory must remain outside the scope of this paper which retains a more traditional field approach, albeit with GA (Geometric Algebra) [**28,29**].

A way around the main difficulty referred to above of a homogeneous 3-d time was given by Cole and Buchanan who showed that changing the temporal vector required so much energy that observed time essentially had a single direction [**30**]. The 3-d time would not then be homogeneous. In this present paper, conventional energy that can be exchanged is linked, as usual, to a single reciprocal principal time along $Ot_3$. Two additional dimensions of time form a preferred 'transverse' temporal plane $Ot_1t_2$ lying normal to the direction of 'principal' time. The reciprocal times in these transverse dimensions are linked to mass energy and are not allowed to be rotated into free energy without energy exchange, that is not discussed here. A six-dimensional wave equation now becomes a Klein-Gordon equation:

$$-(\partial_{s1}^2 + \partial_{s2}^2 + \partial_{s3}^2)\Phi + M_o^2\Phi + \partial_{t3}^2\Phi = 0 ;  \quad\quad\quad (1.1)$$

where $M_o^2\Phi = (\partial_{t1}^2 + \partial_{t2}^2)\Phi$ ; $(c = \hbar = 1)$.

Spatial homogeneity permits any re-orientation of the spatial axes $\{Os_n\}$, but temporal space will not be allowed to have this homogeneity. The axes, $Ot_3$, of 'principal' or 'observed' time, always remains normal to the temporal plane $Ot_1t_2$. When changing to a frame of reference moving with velocity ***v***, the spatial axes may be rotated so that the velocity lies along $Os_3$ but the temporal direction must always lie along $Ot_3$. The axes $Ot_1$, $Ot_2$ as well as $Os_1$ and $Os_2$ are all transverse to ***v***. A classic 'Lorentz' transformation in the '3+1' spacetime that is embedded in the '3+3' spacetime can be formed. The transverse components both of space and of time remain unaltered (Appendix C). The 'transverse' temporal operator $M_o^2$ in equation (1) is then invariant under Lorentz transformations, allowing $M_o$ to be identified as a (normalised) *rest* mass. The (free energy)$^2$ is identified by $k^2 = -\partial_{t1}^3$ with the opposite sign to (mass energy)$^2$ where $M_o^2 = (\partial_{t1}^2 + \partial_{t2}^2)$. As already indicated, rotations between O$t_3$ and O$t_1$ or O$t_2$ can only be allowed if there is some (yet to be discussed) process of mass energy and free energy exchange.

Another major difficulty is that while there are elegant theories for Maxwell's equations with 3-d time, there is no compelling evidence for any of the variants. Cole has 3-component potentials **A** and **Φ** [**31**, **32**]. These lead to a 9-component **E** field in a 3 × 3 spacetime array along with a new 3-component 'magnetic' field **W** augmenting the usual 3-component **B** field. Dattoli & Mignani and also Vysin have fields similar to **W** but with only |**Φ**| and |**A**| being measurable [**33**, **34**]. Teli and Teli & Palaskar link each dimension of time directly with a dimension in space, giving a pleasingly symmetric form of Maxwell's equations [**35**, **36**]. Barashenkov & Yurlev have similarities with Cole's approach and have causality limiting the solutions to those with positive energies [**37**] . In all these papers, Maxwell's equations are reached in some limit but are not derivations from first principles.

Here a fresh start is made using Geometric Algebra (GA): *the* algebra of multi-dimensional symmetry. For nearly four decades, GA has been brilliantly promoted by Hestenes [**28**, **38 39 40** ] along with many others as outlined by Doran and Lasenby [**29**]. Spacetime Algebra (STA) is the name reserved for a branch of GA that specialises in '3+1' spacetime and this allows Maxwell's equations to be expressed in a single equation (Appendix A). The extension of this work to '3+3' spacetime gives several new results:

(i) Just as the complex derivative of an anti-holomorphic function vanishes [(d/dz) $f(z^*)$ = 0], so in '3+3' spacetime a first order derivative has to vanish for certain multi-vector functions leading to first order differential spacetime equations (section 2).

(ii) Transverse time allows the 3+3 fields to appear as complex fields in 3+1 spacetime but the 'imaginary' *i* is now a real bivector in transverse time (section 3)

(iii) Maxwell's equations in an embedded '3+1' spacetime are *derived* from the vanishing multi-vector derivative in (i). It is then entirely the symmetry of spacetime with zero rest mass that determines Maxwell's equations *including* charges (section 4).

(iv) Real fields and Lorentz invariant charges can be found through a contour integration in transverse time. Positive and negative charges may be linked to poles and zeros (section 5). The standard normal modes are discussed in section (6).

(v) Complex conjugation and space-reversal leads to new solutions in '3+3' spacetime lying on the 'advanced' branch of the Lorentz transformation in '3+1' spacetime with advanced waves [**41**]. However these new solutions have 'negative' energy and give *left-handed* sets of **E**, **B**, and **k** for plane waves (section 7)

(vi) Resonant wave-packets can be formed between right- and left handed waves moving together in space and time at the velocity of light with resonant frequencies characteristic of photons (section 8). Such packets will be shown not to violate causality. Section 9 reaches conclusions and looks ahead to future work.

**2. '3+3' spacetime**

In the hope of increasing accessibility, the work here uses only first principles of GA, such as the commutation rules, and retains much of the language of the vector algebra of Gibbs. This section starts by outlining the notation used here for multi-vectors.

Temporal space has unit orthogonal 1-vectors $\boldsymbol{\gamma_t} = \{\gamma_{t1}, \gamma_{t2}, \gamma_{t3}\}$, where the principal or observed time is in the direction of $\gamma_{t3}$. The spatial unit orthogonal 1-vectors are $\boldsymbol{\gamma_s} = \{\gamma_{s1}, \gamma_{s2}, \gamma_{s3}\}$. With $c = 1$ units used throughout this paper, the co-ordinate vector of an event in spacetime is given from $\mathbf{X} = \{\mathbf{t}, \mathbf{x}\}$ with $\mathbf{t} = (t_1, t_2, t_3)$ and $\mathbf{x} = (x_1, x_2, x_3)$ so that a spacetime 1-vector is given from:

$$\underline{\mathbf{X}} = \boldsymbol{\gamma_s} \cdot \mathbf{x} + \boldsymbol{\gamma_t} \cdot \mathbf{t} = \sum_n \gamma_{sn} x_n + \sum_m \gamma_{tm} t_m \qquad (2.1)$$

Bold roman type indicates 3-component vectors (in space or time as context will show). Underlined bold symbols indicate a multi-vector: here $\underline{\mathbf{X}}$ is a 1-vector which means it has only linear combinations of $\gamma_\mu$ with $\mu$ taking values *s1 - s3* and *t1 - t3*. It can be helpful to regard $\gamma_\mu$ as anti-commuting matrices. Anti-commutation is the key requirement for orthogonality in GA. A metric with 3+3 signature given from $\underline{\mathbf{X}}\,\underline{\mathbf{X}} = \mathbf{t}.\mathbf{t} - \mathbf{x}.\mathbf{x}$ requires commutation /multiplication rules:

$$\gamma_{tm}\gamma_{tn} = -\gamma_{tn}\gamma_{tm}, \; \gamma_{sm}\gamma_{sn} = -\gamma_{sn}\gamma_{sm} \; (m \neq n);$$

$$\gamma_{tm}\gamma_{sn} = -\gamma_{sn}\gamma_{tm}, \; \gamma_{tn}^2 = 1, \; \gamma_{sm}^2 = -1, \text{(all m, n)}; \qquad (2.2)$$

It is clear that '3+3' spacetime has more degrees of freedom and a richer structure than '3+1' spacetime. The 1-vector $\underline{\mathbf{X}}$ in equation (2.1) is spanned by six basic 1-vectors. If two arbitrary 1-vectors, $\underline{\mathbf{X}}_a$ and $\underline{\mathbf{X}}_b$ are formed as in equation (2.1) then a geometric product $\underline{\mathbf{X}}_a \underline{\mathbf{X}}_b$ may be evaluated using equation (3) (see a simpler example in the appendix A, equation A.3). The most general product $\underline{\mathbf{X}}_a \underline{\mathbf{X}}_b$ contains one scalar term (formed from terms $\gamma_\mu \gamma_\mu = +/-1$) and fifteen different types of bi-vector (or 2-vector) spanned by pairs $(\gamma_{tm} \gamma_{tn})$, $(\gamma_{sm} \gamma_{sn})$, $(\gamma_{tm} \gamma_{sn})$ (m ≠ n, m,n taking values 1-3). Further products such as $(\underline{\mathbf{X}}_a \underline{\mathbf{X}}_b \underline{\mathbf{X}}_c)$ lead to twenty different types of 3-vector spanned by $(\gamma_{tm} \gamma_{tn} \gamma_{so})$ and $(\gamma_{sm} \gamma_{sn} \gamma_{to})$ (m,n and o all different from 1-3) along with two special 3-vectors:

$$\varphi_s = (\gamma_{s1} \gamma_{s2} \gamma_{s3}), \; \varphi_t = (\gamma_{t1} \gamma_{t2} \gamma_{t3}) \; ; \; [\text{with } \varphi_s^2 = 1 \; ; \; \varphi_t^2 = -1] \tag{2.3}$$

The order in the products in $\varphi_s$ and $\varphi_t$, once chosen, is preserved to retain the spatial and temporal symmetry. The '3+3' spacetime pseudo scalar $S$ is the ordered product (of all the basis orthogonal 1-vectors) taken here as

$$S = (\gamma_{t1} \gamma_{t2} \gamma_{t3})(\gamma_{s1} \gamma_{s2} \gamma_{s3}) = \varphi_t \varphi_s. \quad [S^2 = 1]$$
(2.4)

Further geometric products lead to fifteen pseudo-bi-vectors (4-vectors) spanned by $(S\gamma_{tm} \gamma_{tn})$, $(S \gamma_{tm} \gamma_{sn})$ and $(S \gamma_{sm} \gamma_{sn})$ (m ≠ n, m,n, taking values 1-3) and six pseudo 1-vectors (5-vectors) spanned by $(S\gamma_{tm})$ and $(S \gamma_{sm})$. There is only one pseudo-scalar $S$. Any pseudo-3-vector remains a 3-vector so that one can consider ten 3-vectors and their ten pseudo-3-vector counterparts as forming the twenty 3-vectors. The most general multi-vector now has sixty-four elements as opposed to sixteen elements in '3+1' STA. This general multi-vector in 3+3 spacetime may be written in the form:

$$\underline{\mathbf{V}} = \{\Psi + S \Psi'\} + \{\gamma_\mathbf{s} \cdot (\mathbf{P}_s + S \mathbf{P}_s') + \gamma_\mathbf{t} \cdot (\mathbf{Q}_t + S \mathbf{Q}_t')\}$$
$$+ \{\gamma_\mathbf{s} \cdot \gamma_\mathbf{t} \cdot (\mathbf{F}_{st} + S \mathbf{F}_{st}') + \varphi_t \gamma_\mathbf{t} \cdot (\mathbf{G}_t + S \mathbf{G}_t') + \varphi_s \gamma_\mathbf{s} \cdot (\mathbf{G}_s + S \mathbf{G}_s')\}$$
$$+ \{\varphi_t (L + SL') + \varphi_t \gamma_\mathbf{s} \cdot \gamma_\mathbf{t} \cdot (\mathbf{H}_{st} + S \mathbf{H}_{st}')\} \tag{2.5}$$

Notes: (a) Bold faced terms such as $\mathbf{P}_s$, $\mathbf{Q}_t$ etc each contain 3 terms

e.g. $\gamma_\mathbf{s} \cdot \mathbf{P}_s = \gamma_{s1} P_{s1} + \gamma_{s2} P_{s2} + \gamma_{s3} P_{s3}$ or $\varphi_t \gamma_\mathbf{t} \cdot \mathbf{G}_t = \varphi_t \gamma_{t1} G_{t1} + \varphi_t \gamma_{t2} P_{t2} + \varphi_t \gamma_{t3} P_{t3}$

(b) Bold faced terms such as $\mathbf{F}_{st}$, $\mathbf{H}_{st}$ each contain 9 terms

e.g. $\gamma_\mathbf{s} \cdot \gamma_\mathbf{t} \cdot \mathbf{F}_{st} = \Sigma_{m=1-3, n=1-3} \gamma_{sm} \gamma_{tn} F_{sm\,tn}$

(c) $\mathbf{P}_s$ gives real components of the spatial 1-vector $\gamma_\mathbf{s} \cdot \mathbf{P}_s$ while $\mathbf{P}_s'$ are real components of the spatial pseudo-1-vector $\gamma_\mathbf{s} \cdot S \mathbf{P}_s'$. Similar remarks apply to all other components.

(d) If a function $f(\underline{\mathbf{X}})$ can be expanded in a convergent power series $\Sigma\{\alpha_o+\alpha_1 \underline{\mathbf{X}} + \alpha_2 \underline{\mathbf{X}} \underline{\mathbf{X}} + \alpha_3 \underline{\mathbf{X}} \underline{\mathbf{X}} \underline{\mathbf{X}} + ...\}$ then this must give a multi-vector of the general form $\underline{\mathbf{V}}$ as in equation (2.5).

(e) Equation (2.5) is not unique: it could have different ordering and different signs but once chosen these must be carried through consistently. Once chosen, the order of symbols matters e.g. $\gamma_s .(S\, \mathbf{P}'_s) = - (S\, \mathbf{P}'_s).\gamma_s$.

The next step is to discuss derivatives of multi-vectors. However before looking at these recall the derivative for a complex variable $w = u + i\, v$ with $w^* = u - i\, v$ :

$$\partial/\partial\, w = \tfrac{1}{2}(\partial/\partial\, u - i\, \partial/\partial\, v)\,;\ \ \partial/\partial\, w^* = \tfrac{1}{2}(\partial/\partial\, u + i\, \partial/\partial\, v) \tag{2.6}$$

This yields for any function $f(w^*)$

$$(\partial/\partial\, w)\, f(w^*) = 0\,. \tag{2.7}$$

A related result is found in 3+3 spacetime. The 3+3 spacetime derivative can be formed using: $\boldsymbol{\partial}_t = (\partial_{t1}, \partial_{t2}, \partial_{t3})$, $\boldsymbol{\partial}_s = (\partial_{s1}, \partial_{s2}, \partial_{s3})$ to give a differential GA operator:

$$\underline{\mathbf{D}} = (\boldsymbol{\partial}_t.\boldsymbol{\gamma}_t - \boldsymbol{\partial}_s.\boldsymbol{\gamma}_s)\,; \tag{2.8}$$

The sign of $.\boldsymbol{\gamma}_s$ reflects the reciprocal basis required for describing the covariant derivative as opposed to the contravariant position vector in equation (2.1). The $\gamma_\mu$ are independent of spacetime in this work so that $\partial_\mu \gamma_\mu = \gamma_\mu \partial_\mu$. Note that $\underline{\mathbf{D}}\,\underline{\mathbf{X}} = 6$ appropriate to a '3+3' dimensional space. The equivalent of equation (2.7) requires a modified $\underline{\mathbf{X}}$ such that

$$\underline{\mathbf{X}}^+ = (\mathbf{t}.\boldsymbol{\gamma}_t - \mathbf{x}.\boldsymbol{\gamma}_s)\,;\ \ \underline{\mathbf{D}}\,\underline{\mathbf{X}}^+ = 0\,; \tag{2.9}$$

Writing the general multi-vector $\underline{\mathbf{V}}$ in equation (2.5) as $f(\underline{\mathbf{X}})$, then $f(\underline{\mathbf{X}}^+)$ gives a multi-vector $\underline{\mathbf{V}}^+$. The required change of sign of $\boldsymbol{\gamma}_s$ in equation (2.5) is accomplished by a transformation $\Pi = \gamma_{t1}\gamma_{t2}\gamma_{t3}$ with $\Pi^{-1} = \gamma_{t3}\gamma_{t2}\gamma_{t1}$ giving:

$$\underline{\mathbf{X}}^+ = \Pi\,\underline{\mathbf{X}}\,\Pi^{-1}\,;\ \ (\underline{\mathbf{X}}^+)^N = \Pi\,(\underline{\mathbf{X}})^N\,\Pi^{-1}\,;\ \ \Pi f(\underline{\mathbf{X}})\,\Pi^{-1} = f(\underline{\mathbf{X}}^+) = \Pi\,\underline{\mathbf{V}}\,\Pi^{-1} = \underline{\mathbf{V}}^+\,. \tag{2.10}$$

Unfortunately, because the $\gamma_\mu$ do not commute, leading to $\underline{\mathbf{X}}^+\underline{\mathbf{X}}^+ = \underline{\mathbf{X}}\,\underline{\mathbf{X}}$ then only a restricted set of functions $f(\underline{\mathbf{X}}^+)$ have the correct symmetry that $\underline{\mathbf{D}}\,f(\underline{\mathbf{X}}^+) = 0$. In particular if $\underline{\mathbf{X}}$ lies on a generalised light cone where $\underline{\mathbf{X}}\,\underline{\mathbf{X}} = 0$ (or a constant) then a wide range of $f(\underline{\mathbf{X}}^+)$ can be constructed so as to satisfy $\underline{\mathbf{D}} f(\underline{\mathbf{X}}^+) = 0$. Consequently multivectors $\underline{\mathbf{V}}^+$ are sought that have the required symmetry such that :

$$\underline{\mathbf{D}}\,\underline{\mathbf{V}}^+ = 0. \tag{2.11}$$

Now split $\underline{\mathbf{V}}^+$ into $(\underline{\mathbf{V}}^+_e + \underline{\mathbf{V}}^+_o)$ where subscripts e/o imply even and odd 'grades' of multi-vector. Scalars, bi-vectors, 4-vectors and the pseudo-scalar are the even grades while 1-, 3-, 5- vectors are the odd grades. Even-grade multi-vectors multiply together to give even-grade multi-vectors and form a sub-algebra. This even sub-algebra has been termed as the spinor algebra by Hestenes [**42**]. If $\underline{\mathbf{V}}_e = f_e(\underline{\mathbf{X}}^+)$ then it will follow $f_e$ has to be an even function so that $f_e(\underline{\mathbf{X}}^+) = f_e(-\underline{\mathbf{X}}^+)$. It will be these terms that through $\underline{\mathbf{D}}\,\underline{\mathbf{V}}^+_e = 0$ will lead to the Dirac equation in a later paper.

In this paper we concentrate on the odd-grade multi-vectors $\underline{\mathbf{V}}^+_o$ that will require $\underline{\mathbf{V}}_o = f_o(\underline{\mathbf{X}}^+)$ where $f_o(\underline{\mathbf{X}}^+) = -f_o(-\underline{\mathbf{X}}^+)$. Selection the odd terms from equation (2.5) modified as in equation (2.10):

$$\underline{\mathbf{V}}^+_o = \{-\boldsymbol{\gamma}_s \cdot (\mathbf{P}_s - S\,\mathbf{P}_s^/) + \boldsymbol{\gamma}_t \cdot (\mathbf{Q}_t - S\,\mathbf{Q}_t^/)\} + \{\varphi_t\,(L - S\,L^/) - \varphi_t\,\boldsymbol{\gamma}_s \cdot \boldsymbol{\gamma}_t \cdot (\mathbf{H}_{st} - S\,\mathbf{H}_{st}^/)\} \tag{2.12}$$

The equivalent of equation (2.11) then becomes

$$\underline{\mathbf{D}}\,\underline{\mathbf{V}}^+_o = 0 \tag{2.13}$$

This will shortly be shown to give Maxwell's equations along with sources .

## 3. Transverse time

This section shows how a preferred transverse plane, spanned by 1-vectors $\gamma_{t1}$ and $\gamma_{t2}$, can create a '3+1' 'complex' spacetime embedded in the '3+3' spacetime. The 'direction' of principal time is determined by the 1-vector $\gamma_{t3}$. The bi-vector $\gamma_{t1}\gamma_{t2} = i$ defines $i$ with $i^2 = -1$ and $\gamma_{t1} i = \gamma_{t2}$. Now consider a temporal 1-vector $\underline{\mathbf{A}}_t = \boldsymbol{\gamma}_t \cdot \mathbf{A}_t$ where $\mathbf{A}_t = \{a_{t1}, a_{t2}, a_{t3}\}$:

$$\boldsymbol{\gamma}_t \cdot \mathbf{A}_t = \gamma_{t3}\,A_{t3} + \gamma_{t1}\,A_T = \gamma_{t3}\,A_{t3} + A_T^*\,\gamma_{t1} \tag{3.1}$$

$$A_T = (a_{t1} + i\,a_{t2})\,;\quad A_T^* = (a_{t1} - i\,a_{t2}). \tag{3.2}$$

Note that $\gamma_{t1} A_T \neq A_T \gamma_{t1}$ so that ordering remains important. However the bivector $i$ commutes with each of the '3+1' spacetime 1-vectors ( $\gamma_{s1}, \gamma_{s2}, \gamma_{s3}$ and $\gamma_{t3}$) just like the classic imaginary scalar $i$.

The temporal gradient operator $\partial_t \cdot \gamma_t$ can now be written as:

$$\partial_t \cdot \gamma_t = \gamma_t \cdot \partial_t = \nabla_T \gamma_{t1} + \partial_{t3} \gamma_{t3} = \gamma_{t1} \nabla_T^* + \gamma_{t3} \partial_{t3}.$$
(3.3)

where $\nabla_T = (\partial_{t1} - i \partial_{t2})$ ; $\nabla_T^* = (\partial_{t1} + i \partial_{t2})$. Note that $\nabla_T (t_1 + i\, t_2) = 2$ (the number of transverse temporal dimensions). The differential operator $\underline{D}$ of equation (2.8) can then be written in the forms:

$$\underline{D} = (-\partial_s \cdot \gamma_s + \nabla_T \gamma_{t1} + \partial_{t3} \gamma_{t3}) = (-\partial_s \cdot \gamma_s + \gamma_{t1} \nabla_{\gamma T}^* + \partial_{t3} \gamma_{t3})$$
(3.4)

Setting $\nabla_T^* \nabla_T = \nabla_T \nabla_T^* = M_0^2$, a wave-equation turns into a Klein-Gordon equation:

$$\underline{D}\,\underline{D}\,\Phi = -\partial_s \cdot \partial_s \Phi + \partial_{t3} \partial_{t3} \Phi + M_0^2 \Phi = 0 \qquad (3.5)$$

As already noted, non-zero mass is to be considered in a later paper discussing the Dirac equation that can be found from the even grade multi-vectors: $\underline{D}\, \underline{V}^+_e = 0$. In this present paper, zero 'mass' ($M_0 = 0$) is assumed for this wave-equation (3.5) in '3+3' space. The result is equivalent to the classic wave-equation in '3+1' space. Non-trivial fields with $M_0 = 0$ can be found, for example, from harmonic functions $f(t_1 + i\, t_2)$ where $\nabla_T^* \nabla_T f = 0$ but $\nabla_T f \neq 0$.

The notation in equation (2.12) for the multi-vector $\underline{V}_0^+$ is now changed to make use of $i$ as in equation (3.2). Initially define:

$$I = \gamma_{t3} \gamma_{s1} \gamma_{s2} \gamma_{s3} = \gamma_{t3} \varphi_s \qquad (3.6)$$

Here $I$ is the pseudo-scalar for the embedded '3+1' spacetime as used in STA (appendix A). The '3+3' spacetime pseudo-scalar is $S = i\, I$. After a little algebra one may write:

$$\underline{V}_o^+ = \gamma_s \cdot \mathbf{K}_s + \gamma_{t3} U + \gamma_{t1} V_T + \gamma_{t1} \gamma_s \cdot \gamma_{t3} \mathbf{W}_{sT} \qquad (3.7)$$

where $\mathbf{K}_s = \{ K_{s1}, K_{s2}, K_{s3} \}$; $\mathbf{W}_{sT} = \{ W_{s1T}, W_{s2T}, W_{s3T} \}$ with the general form
$K_{sn} = -[P_{sn} + i\, H_{snt3} - I\, (H_{snt3}' + i\, P_{sn}')]$ ; $W_{snT} = -(+i\, H_{snT} + I\, H_{snT}')$ ; (n=1-3) ;
$V_T = (Q_T - S\, Q_T')$ ; $U = [(Q_{t3} + i\, L) - I\, (L' - i Q_{t3}')]$

The subscript $_T$ implies complex terms like $A_T$ as in equation (3.2). The primed terms (e.g. $Q_T{'}$) have no necessary relationship with unprimed terms (e.g. $Q_T$). As before, the ordering matters (e.g. $\gamma_s \cdot K_s \neq K_s \cdot \gamma_s$). However no significance should be attached to particular signs of the components at this stage so long as they are retained consistently in subsequent developments.

## 4. Maxwell's equations

Taking the cue from the methods of STA (Appendix A), start with an *odd* grade multi-vector $\underline{V}_o{}^+ = f_o(X^+)$, [equation (3.7)] and using equation (2.13):

$$\underline{D}\,\underline{V}^+{}_o = (-\partial_s \cdot \gamma_s + \nabla_T \gamma_{t1} + \partial_{t3}\gamma_{t3})(\gamma_s \cdot K_s + \gamma_{t3} U + \gamma_{t1} V_T + \gamma_{t1}\gamma_s \cdot \gamma_{t3} W_{sT}) = 0.$$
(4.1)

This is evaluated using the commutation rules of equations (2.2) and the identity $\varphi_s = \gamma_{t3} I$:

$$\underline{D}\,\underline{V}^+{}_o = [\partial_s \cdot K_s + \partial_{t3} U + \nabla_T V_T] + \gamma_s \cdot \gamma_{t3} [(\partial_s \times I\, K_s) - \partial_s U - \partial_{t3} K_s + \nabla_T W_{sT}]$$
$$+ \gamma_{t1}\gamma_{t3} [\nabla_T{}^* U - \partial_s \cdot W_{sT} - \partial_{t3} V_T]$$
$$+ \gamma_{t1}\gamma_s \cdot [\nabla_T{}^* K_s + (\partial_s \times I\, W_{sT}) + \partial_s V_T + \partial_{t3} W_{sT}] = 0 \qquad (4.2)$$

Each bracketed term [--] in equation (4.2) contains no multi-vectors that are common to any of the other [--] terms and each [--] is therefore zero, leading to four equations. Pre-operate on the last two bracketed terms in equation (4.2) by the operator $(\nabla_T \gamma_{t1})$ to arrive at four separate complex multi-vector equations *demanded only by symmetry*:

$$[\partial_s \cdot K_s + \partial_{t3} U + \nabla_T V_T] = 0 \qquad (4.3)$$

$$\gamma_s \cdot \gamma_{t3} [(\partial_s \times I\, K_s) - \partial_s U - \partial_{t3} K_s + \nabla_T W_{sT}] = 0 \qquad (4.4)$$

$$\gamma_{t3} [\nabla_T \nabla_T{}^* U - \partial_s \cdot \nabla_T W_{sT} - \partial_{t3} \nabla_T V_T] = 0 \qquad (4.5)$$

$$\gamma_s \cdot [\nabla_T \nabla_T{}^* K_s + (\partial_s \times I\, \nabla_{\gamma T} W_{sT}) + \partial_s \nabla_T V_T + \partial_{t3} \nabla_T W_{sT}] = 0 \qquad (4.6)$$

Before manipulating this into the classic Maxwell form, note that it is possible to have

$$\nabla_T{}^* \nabla_T V_T = 0 \; ; \; \nabla_T{}^* \nabla_T W_{sT} = 0 \qquad (4.7)$$

$$\nabla_T \nabla_T{}^* K_s = k_s \neq 0 \; ; \; \nabla_T \nabla_T{}^* U = u \neq 0 \; . \qquad (4.8)$$

Equation (4.8) does not contradict equation (4.7) provided that $\nabla_T \nabla_T{}^*$ operating on equations (4.3) and (4.4) gives non-trivial solutions for $k_s$ and $u$. Because of equation (4.7) these non-trivial solutions require:

$\partial_s \cdot \mathbf{k_s} + \partial_{t3} u = 0$ ; $\gamma_s \cdot \gamma_{t3} [(\partial_s \times I \mathbf{k_s}) - \partial_s u - \partial_{t3}\mathbf{k_s}] = 0$. (4.9)

Now $I$ commutes with $\gamma_s \cdot \gamma_{t3}$ and consequently equations (4.9) can be shown to require that the spatial and principal time components in $u$ and $\mathbf{k_s}$ satisfy the '3+1' wave equation.

$\nabla_s^2 \mathbf{k_s} - \partial_{t3}\partial_{t3} \mathbf{k_s} = 0$ ; $\nabla_s^2 u - \partial_{t3}\partial_{t3} u = 0$ (4.10)

Now if $u$ and $\mathbf{k_s}$ were the only terms helping to form U and $\mathbf{K_s}$ then equations (4.3) and (4.4) would not hold for non zero $\nabla_T V_T$ and $\nabla_T W_{sT}$. Consequently $\mathbf{K_s}$ and U must be composed of two parts: one part satisfies equation (4.10) and the other part satisfies $\nabla_T \nabla_T^* \mathbf{K_s} = 0$ ; $\nabla_T \nabla_T^* U = 0$. All parts will satisfy the massless wave-equation in '3+1' spacetime except possibly, as will be shown later, at isolated singularities in transverse time.

Now define a complex bivector $\mathbf{F_s}$:

$\gamma_s \cdot \gamma_{t3} \mathbf{F_s} = \gamma_s \cdot \gamma_{t3} [-\nabla_T \mathbf{W_{sT}}]$ (4.11)

and using equation (4.8) along with (4.5) and (4.6), the excitation of the fields $\mathbf{F_s}$ is determined from:

$\gamma_{t3} [\partial_s \cdot \mathbf{F_s} - (\rho + S\rho')] = 0$ (4.12)

$\gamma_s \cdot [(\partial_s \times iS \mathbf{F_s}) - (\mathbf{J} + S\mathbf{J'}) - \partial_{t3} \mathbf{F_s}] = 0$ (4.13)

where

$\rho + S\rho' = -(u - \partial_{t3} \nabla_T V_T)$ ; $\mathbf{J} + S\mathbf{J'} = -(\mathbf{k_s} + \partial_s \nabla_T V_T)$ (4.14)

$\partial_{t3} \rho + \partial_s \cdot \mathbf{J} = 0$ ; $\partial_{t3} \rho' + \partial_s \cdot \mathbf{J'} = 0$ (4.15)

The quantities $\rho$ and $\mathbf{J}$ (or $\rho'$ and $\mathbf{J'}$) appear like charge and current with the usual continuity relationship.

The fields $\mathbf{F_s}$ in equation (4.13) are composed of bi-vectors and 4-vectors and may be written as:

$\gamma_s \cdot \gamma_{t3} \mathbf{F_s} = \gamma_s \cdot \gamma_{t3} [\mathbf{F_{s1}} + i \mathbf{F_{s2}} - S i (\mathbf{F_{s3}} + i \mathbf{F_{s4}})]$ (4.16)

The signs are not really significant provided that they are consistently maintained. There are distinctive symmetries. The fields $\gamma_s \cdot \gamma_{t3} (\mathbf{F_{s1}} + i \mathbf{F_{s2}})$ are 'complex space-time' bi-vectors while the fields $\gamma_s \cdot \gamma_{t3} [S i (\mathbf{F_{s3}} + i \mathbf{F_{s4}})]$ are 'complex space-space' bi-vectors. Because the equations (4.12) and (4.13) are linear, one can in the first instance set the pseudo-sources $S\rho^/$ and $SJ^/$ to zero and consider only the excitations ρ and **J**.

$$\gamma_{t3} [\partial_s \cdot \mathbf{F_s} - \rho ] = 0 \qquad (4.17)$$

$$\gamma_s \cdot [ (\partial_s \times iS\, \mathbf{F_s}) - \mathbf{J} - \partial_{t3} \mathbf{F_s}] = 0 \qquad (4.18)$$

Define new bivector fields $\gamma_s \cdot \gamma_{t3}\, \mathbf{E}$ and $\gamma_s \cdot \gamma_{t3}\, I\, \mathbf{B}$ from:

$$\gamma_s \cdot \gamma_{t3}\, \mathbf{E} = \gamma_s \cdot \gamma_{t3}\, (\mathbf{F_{s1}} + i\, \mathbf{F_{s2}}) \quad \gamma_s \cdot \gamma_{t3}\, I\, \mathbf{B} = -\gamma_s \cdot \gamma_{t3}\, i\, S\, (\mathbf{F_{s3}} + i\, \mathbf{F_{s4}}) \qquad (4.19)$$

The **E**-fields are the complex space-time bi-vectors, while the **B**-fields are the complex space-space bi-vectors. Equating the different grades of multi-vector in equations (4.17) and (4.18) produces the 'Maxwell equations':

$$\partial_s \cdot \mathbf{E} = \rho \quad ; \quad \partial_s \cdot \mathbf{B} = 0 \quad ; \qquad (4.20)$$

$$\partial_s \times \mathbf{E} = -\partial_{t3} \mathbf{B} \quad ; \quad \partial_s \times \mathbf{B} = \partial_{t3} \mathbf{E} + \mathbf{J} \quad ; \qquad (4.21)$$

Apart from the fact that all the terms are complex, these equations mirror exactly the classic Maxwell equations along with their sources [**43**].

Now consider non-zero pseudo-sources $S\rho^/$ and $SJ^/$ but set the excitations ρ and **J** to zero. These pseudo-sources will excite $\mathbf{F}^/_s$ where one may write:

$$\gamma_s \cdot \gamma_{t3}\, S\, \mathbf{F}^/_s = \gamma_s \cdot \gamma_{t3}\, [-i\, S\, (-\mathbf{F}^/_{s2} + i\, \mathbf{F}^/_{s1}) + (\mathbf{F}^/_{s4} - i\, \mathbf{F}^/_{s3})]$$

$$= \gamma_s \cdot \gamma_{t3}\, [I\, \mathbf{B}^/ + \mathbf{E}^/ ] \qquad (4.22)$$

The fields $I\, \mathbf{B}^/$ and $\mathbf{E}^/$ have essentially the same spacetime symmetries as $I\, \mathbf{B}$ and $\mathbf{E}$ and after a little manipulation one find that $I\, \mathbf{B}^/$ and $\mathbf{E}^/$ satisfy equations (4.20) and (4.21) with primed quantities replacing unprimed quantities. The pseudo-fields in the 3+3 spacetime behave the same as the non-pseudo fields. It follows that there are no sources for either div **B** or div $\mathbf{B}^/$ in 3+1 spacetime: i.e. according to this theory, symmetry demands that there are no magnetic monopoles.

The remaining unexplained equations (4.3) and (4.4) can now be given a classical significance by writing

$$\mathbf{A_s} = \mathbf{K_s} + \partial_s \psi \quad ; \quad A_t = U - \partial_{t3} \psi \qquad (4.23)$$

$$\partial_s \cdot \partial_s \psi - \partial_{t3}^2 \psi = \nabla_T V_T \quad ; \qquad (4.24)$$

leading to the interpretation of vector and scalar potentials $\mathbf{A_s}$ and $A_t$ that define scalar and bi-vector fields as follows:

$$\partial_s \cdot \mathbf{A_s} + \partial_{t3} A_t = 0 ; \tag{4.25}$$

$$\gamma_s \cdot \gamma_{t3} [(\partial_s \times I \mathbf{A_s}) - \partial_s A_t - \partial_{t3} \mathbf{A_s} - (\mathbf{E} + I \mathbf{B})] = 0 \tag{4.26}$$

Equation (4.25) is usually called the Lorentz condition. Separating out the multi-vector components in equation (4.26) yields scalar and vector potentials for the $\mathbf{E}$ and $\mathbf{B}$ fields:

$$(\partial_s \times \mathbf{A_s}) = \mathbf{B} \qquad -\partial_s A_t - \partial_{t3} \mathbf{A_s} = \mathbf{E} \tag{4.27}$$

The classic form Maxwell's equations in '3+1' space time along with charges and potentials have now been *derived* from a 1$^{st}$ order derivative that has to be zero in the '3+3' spacetime fields as demanded by symmetry. The zero divergence of the B-fields is a matter of spacetime symmetry suggesting that '3+3' spacetime symmetry demands the non-existence of magnetic mono-poles in '3+1' spacetime. However at present one cannot compare theory and experiment because these fields and sources are complex functions of $t_1 + it_2$. A reality check has to be made!

**5. Real fields and charge**

Because sources and pseudo-sources give the same types of field, only the sources $\rho$ and $\mathbf{J}$ are considered. To move towards real solutions, write $\tau = t_1 + i\, t_2$ and suppose that

$$\rho = \Xi(\tau) v(x_1, x_2, x_3, t_3) \text{ with } \Xi(\tau) = [-1/(2\pi i)] \log[q(\tau)] \tag{5.1}$$

where $v(x_1, x_2, x_3, t_3)$ is real; $q(\tau)$ has simple zeros at $\tau = (t_{1n} + i\, t_{2n})$ and simple poles at $\tau = (t_{1p} + i\, t_{2p})$. From the principle of the argument [**44**]

$$\oint \Xi(\tau) \, d\tau = (P - N)\, v(x_1, x_2, x_3, t_3). \tag{5.2}$$

where N = number of zeros and P = number of poles in $q(\tau)$. The choice of sign in equation (5.1) has been deliberately chosen to suggest that N represents negative 'charge' while P represents positive 'charge' in appropriate normalised units. One suddenly sees a rationale suggesting that positive and negative charge may be related to isolated poles and zeros in a complex plane. The behaviour of both $\rho$ and $\mathbf{J}$ in transverse time is identical but the pair [$\mathbf{J}$, $\rho$] form a classic four-vector in 3+1 spacetime. All transverse temporal terms are Lorentz invariants so that the charges P and N are Lorentz invariant. Taking the conjugate gives a meaning to the term 'charge conjugation' by reversing the sign e.g.:

$$\oint \nabla_T^* V_T^* d\tau^* = (N - P)\, v(x_1, x_2, x_3, t_3). \tag{5.3}$$

A contour integration now gives rise to real fields as if these were generated from 'positive' and 'negative' charges.

A conjecture in this theory is that any real local measurement in '3+1' spacetime takes an average in the form of a closed contour integral in the transverse temporal plane of '3+3' spacetime. This means that the orientation of the transverse temporal axes cannot be determined by classical observations: certainly in accord with experimental observation!

The physics of rotation in transverse time (appendix D) along with positive and negative frequencies provide an integrable field theory in '3+1' space time with all the requirements for a clear arrow of time [**45**]. Real Maxwell's equations, with charge sources and potentials can now be said to be a necessary outcome of the symmetry of embedding a '3+1' spacetime into a '3+3' spacetime. All the fields satisfy the '3+1' wave-equation except for the potentials (related to $\mathbf{K}_s$ and U) which will not satisfy this equation at isolated singularities in transverse time representing the charge sources.

## 6. Plane wave normal modes

Plane waves varying as cos(**k.x**) and sin(**k.x**) provide a start for the treatment of normal modes (see for example Cohen-Tannoudji et. al.[**46**]). Because of the special meaning attached to *i*, care is now needed in forming a complex Fourier analysis. Using the transverse gauge of equations (4.24)-(4.26) write $\boldsymbol{\gamma}_s \cdot \mathbf{A}_\perp(\mathbf{r}, t) = \boldsymbol{\gamma}_s \cdot [\mathbf{A}_\perp{'}(\mathbf{r}, t) + i\, \mathbf{A}_\perp{''}(\mathbf{r}, t)]$ with

$$(\partial_{t3}^2 - \partial_\mathbf{s} \cdot \partial_\mathbf{s})\, \mathbf{A}_\perp = 0\ ;\quad \partial_\mathbf{s} \cdot \mathbf{A}_\perp = 0\ ;$$
(6.1)

For waves with real wave-vectors **k**, a real spatial Fourier analyses using cos(**k.x**) and sin(**k.x**) can be made for each of the real $\mathbf{A}_\perp{'}$ and $\mathbf{A}_\perp{''}$ which then combine to give:

$$\mathbf{A}_\perp (\mathbf{r}, t_3) = \mathbf{A}_\perp{'} + i\, \mathbf{A}_\perp{''} = \Sigma_k\, \mathbf{A}_\perp{\hat{}}(\mathbf{k}, t)\, \exp(i\, \mathbf{k.x})\ ; \tag{6.2}$$

Where the caret gives the k-components of the vector potential. However $\mathbf{A}_\perp{\hat{}}(-\mathbf{k}, t_3)$ only equals $\mathbf{A}_\perp{\hat{}}(\mathbf{k}, t_3)^*$ if one can be assured that $\mathbf{A}_\perp(\mathbf{r}, t_3)$ is real     Never the less, for a given **k**, equation (6.1) shows that $\mathbf{A}_\perp{\hat{}} \cdot \mathbf{k} = 0$ so that $\mathbf{A}_\perp{\hat{}}$ is normal to **k**:

$$A_\perp\!{}^\wedge\,(\mathbf{k}, t_3) = A_\perp\!{}^\wedge\,(\mathbf{k})\,\exp(-/+i\,kt_3)\,;\text{ with } \mathbf{k}.\mathbf{k} = k^2\,;\ k > 0. \tag{6.3}$$

For waves travelling in the forward direction with a phase factor ($\mathbf{k}.\mathbf{x} - kt_3$), there are then fields with either 'positive' or 'negative frequencies: $\mathbf{A}_\perp\!{}^\wedge\,(+/-\mathbf{k})\,\exp[-/+i\,(kt_3 - \mathbf{k}.\mathbf{x})]$. The physicist's convention is adopted with 'positive' frequencies given from $\exp(-i\,kt_3)$. Defining a unit vector $\mathbf{k}^\wedge = \mathbf{k}/k$ then equation (6.3) gives for a positive frequency waves:

$$\boldsymbol{\gamma_s}.\boldsymbol{\gamma_t}[\mathbf{F}(\mathbf{k})_{\mathbf{r}\,t}] = \boldsymbol{\gamma_s}.\boldsymbol{\gamma_t}[\mathbf{E} + I\mathbf{B}] = \boldsymbol{\gamma_s}.\boldsymbol{\gamma_t}[\,ik\{\,\mathbf{A}_\perp\!{}^\wedge\,(\mathbf{k}) + I\,[\mathbf{k}^\wedge\!\times \mathbf{A}_\perp\!{}^\wedge\,(\mathbf{k})]\}\,\exp[-i(kt_3 - \mathbf{k}.\mathbf{x})]\,] \tag{6.4}$$

Even though $\mathbf{E}$ and $\mathbf{B}$ may be complex, $\mathbf{E}.\mathbf{B} = 0$ and $\mathbf{k}^\wedge\!\times \mathbf{E} = \mathbf{B}$ hence retaining the classic right-handed set of vectors $\mathbf{E}$, $\mathbf{B}$ and $\mathbf{k}^\wedge$ in '3+1' spacetime.

## 7. Left-handed fields in '3+3' spacetime

The chirality of any three dimensional system is a statement about whether the system of its three orthogonal axes is right- or left- handed, and simply re-labelling the axes does not give a new solution. SpaceTime Algebra gives a clear answer that, with the sign conventions chosen, the symmetry of space and time forces the $\mathbf{E}$-, $\mathbf{B}$-fields and $\mathbf{k}$ to form a right handed set for plane-waves propagating in the direction $\mathbf{k}$. Are there situations where a genuine physical change can give a left handed chirality for $\mathbf{E}$-, $\mathbf{B}$-fields and $\mathbf{k}$?

Left handed materials were envisaged over 35 years ago by Veselago [**47**]. There is now considerable interest in the physics, the properties and the manufacture of these 'meta-materials' (for example **48**, **49**, **50**). The special left-handed properties occur in meta-materials over a relatively narrow frequency range through manufacturing periodic structures that effectively change the sign of the permeability and/or permittivity in such a way that the e.m. waves have a negative group velocity over that frequency range. These left-handed waves do not co-exist with right-handed waves at the same frequency. With three dimensional time it will be shown that there is a left-handed solution as well as the right handed solution for the Maxwellian waves: both types of wave can in principle co-exist over the full spectrum of frequencies.

One needs to know how/if the chirality of the solution of equation (6.4) can be changed so as to change the right-handed waves into left-handed waves. The key to transformations in GA/STA is to find a 'rotor' $\mathbf{R}$ and its inverse $\mathbf{R}^{-1}$ such that for any multi-vector $\underline{\mathbf{A}}$ the required transformation is given from $\underline{\mathbf{A}}_\mathbf{R} = \mathbf{R}\,\underline{\mathbf{A}}\,\mathbf{R}^{-1}$, analogous to matrix transformations. The transformation of a product is the product of transformations:-

$[\underline{\mathbf{A}}\,\underline{\mathbf{B}}\,\underline{\mathbf{C}}]_\mathbf{R} = (\mathbf{R}\,\underline{\mathbf{A}}\,\mathbf{R}^{-1})(\mathbf{R}\,\underline{\mathbf{B}}\,\mathbf{R}^{-1})(\mathbf{R}\,\underline{\mathbf{C}}\,\mathbf{R}^{-1}) = \mathbf{R}\,[\underline{\mathbf{A}}\,\underline{\mathbf{B}}\,\underline{\mathbf{C}}]\mathbf{R}^{-1}$. In particular $f(\underline{\mathbf{A}})_\mathbf{R} = \mathbf{R}\,f(\underline{\mathbf{A}})\,\mathbf{R}^{-1}$ for well behaved functions $f$. Rotations and Lorentz transformations, discussed further in appendix B, do not change chirality. Define a transformation $\mathbf{R_c}$ that reverses $\gamma_{t2}$ but not $\gamma_{t1}$ or $\gamma_{t3}$. Then $\mathbf{R_c}\,i\,\mathbf{R_c}^{-1} = -\gamma_{t1}\gamma_{t2} = -i$. $\mathbf{R_c}$ also must reverse space ($\gamma_{sn} \to -\gamma_{sn}$):

$\mathbf{R_c}\,\gamma_{tn}\,\mathbf{R_c}^{-1} = \gamma_{tn}$, n = 1,3    $\mathbf{R_c}\,\gamma_{sn}\,\mathbf{R_c}^{-1} = -\gamma_{sn}$  n = 1,2,3 ;   $\mathbf{R_c}\,\gamma_{t2}\,\mathbf{R_c}^{-1} = -\gamma_{t2}$

(7.1)

These transformations are all satisfied if

$\mathbf{R_c} = \gamma_{t2}\varphi_s$ ;  $\mathbf{R_c}^{-1} = \varphi_s\gamma_{t2}$ ,  $\mathbf{R_c}\,\mathbf{R_c}^{-1} = \mathbf{R_c}^{-1}\,\mathbf{R_c} = 1$

(7.2)

Space-reversal changes the spatial chirality while conjugation ($i \to -i$) changes the temporal chirality. However $\mathbf{R_c}\,S\,\mathbf{R_c}^{-1} = S$ so that one still has exactly the same '3+3' pseudo-scalar preserving the *overall* chirality of '3+3' spacetime. Equation (2.13) now becomes:

$\underline{\mathbf{D}}_\mathbf{c}\,\underline{\mathbf{V}}^+{}_{oc} = 0$ (7.3)

where $\underline{\mathbf{D}}_\mathbf{c} = \mathbf{R_c}\,\underline{\mathbf{D}}\,\mathbf{R_c}^{-1}$ and $\underline{\mathbf{V}}^+{}_{oc} = \mathbf{R_c}\,\underline{\mathbf{V}}_o{}^+\,\mathbf{R_c}^{-1}$. Fortunately one does not need to work through the analysis yet again but simply operate on any fields $\mathbf{F}$ to form their conjugate fields $\mathbf{R_c}\,\mathbf{F}\,\mathbf{R_c}^{-1}$.

Using the operator pair $\{\mathbf{R_c}, \mathbf{R_c}^{-1}\}$ as outlined above, additional results are noted.

$\mathbf{R_c}\,\exp(-i\,\theta)\,\mathbf{R_c}^{-1} = \exp(i\,\theta)$ ;   $\mathbf{R_c}\,I\,\mathbf{R_c}^{-1} = -I$ (7.4)

Conjugation then changes fields $\mathbf{F}$ with a 'positive' frequency variation, $\exp[-i(kt - \mathbf{k}.\mathbf{x})]$, into fields $\mathbf{F_c}$ varying with a negative frequency as $\exp[i(kt - \mathbf{k}.\mathbf{x})$ and changes the sign of $I$:

$\mathbf{R_c}\,\gamma_s.\gamma_t\,(\mathbf{E}+I\mathbf{B})\,\mathbf{R_c}^{-1} = -\gamma_s.\gamma_t\,(\mathbf{E_c} - I\mathbf{B_c}) = -\gamma_s.\gamma_t\,[\mathbf{E_c} + I(-\mathbf{B_c})] = -\gamma_s.\gamma_t\,\mathbf{F_c}$ (7.5)

The sign convention means that $\mathbf{E_c}$, $-\mathbf{B_c}$ along with $\mathbf{k}$ now form the right handed set equivalent to $\mathbf{E}$, $\mathbf{B}$ and $\mathbf{k}$. In other words $\mathbf{E_c}$, $\mathbf{B_c}$ with $\mathbf{k}$ form *a left handed* set. The frequency and wave-vector for these conjugate fields has changed sign from the initial fields while keeping the same phase-velocity.

Conjugation changes the source term $\nabla_T V_T$ in equations (4.3) and (4.4) into $\nabla_T^* V_T^*$ but $\nabla_T \nabla_T^* = \nabla_T^* \nabla_T$ so that all the fields remain harmonic functions of transverse time and the zero mass condition is not altered. From section 5 it can therefore be seen that conjugation changes the sign of the charge for the conjugate field equations.

The Lorentz transformation for the fields is discussed in appendix C and is determined by an operator $\mathbf{R_L}$ that operates in a similar manner to $\mathbf{R_c}$. For a boost parameter $\alpha$ along a direction $\mathbf{b}$ in space, the Lorentz 'rotor' and its conjugate are $\mathbf{R_L}$ and $\mathbf{R_{Lc}}$ where:

$\mathbf{R_L} = \exp(½\alpha\, \gamma_{t3}\, \boldsymbol{\gamma_s} \cdot \mathbf{b})$ ; $\mathbf{R_L}^{-1} = \exp(-½\alpha\, \gamma_{t3}\, \boldsymbol{\gamma_s} \cdot \mathbf{b})$;

$\mathbf{R_{Lc}} = \mathbf{R_c}\, \mathbf{R_L}\, \mathbf{R_c}^{-1} = \exp(-½\alpha\, \gamma_{t3}\, \boldsymbol{\gamma_s} \cdot \mathbf{b})$ ; $\mathbf{R_{Lc}}^{-1} = \exp(½\alpha\, \gamma_{t3}\, \boldsymbol{\gamma_s} \cdot \mathbf{b})$  (7.6)

$\mathbf{R_L}$ and $\mathbf{R_{Lc}}$ have opposite values of $\alpha$ indicating they operate on opposite branches of the Lorentz transformation for a given change of velocity. The transformed fields then are $(\boldsymbol{\gamma_s} \cdot \gamma_{t3}\, \mathbf{F})_L$ and $(\boldsymbol{\gamma_s} \cdot \gamma_{t3}\, \mathbf{F_c})_L$ where

$(\boldsymbol{\gamma_s} \cdot \gamma_{t3}\, \mathbf{F})_L = \mathbf{R_L}(\boldsymbol{\gamma_s} \cdot \gamma_{t3}\, \mathbf{F})\mathbf{R_L}^{-1}$ ; $(\boldsymbol{\gamma_s} \cdot \gamma_{t3}\, \mathbf{F_c})_L = \mathbf{R_{Lc}}(\boldsymbol{\gamma_s} \cdot \gamma_{t3}\, \mathbf{F_c})\mathbf{R_{Lc}}^{-1}$
(7.7)

Suppose that it is accepted that the right-handed fields described by $\mathbf{F}$ in equation (6.4) lead to the usual causal waves, excited by some source in the past, carrying energy into the future. These wave can be said to lie on the 'retarded' light cone with 'positive' frequencies. These are traditional waves carrying positive energy that can be linked in quantum theory to their positive frequency. On conjugation of $\mathbf{F}$ into $\mathbf{F_c}$, the frequency changes to be 'negative' but the phase velocity in space is unchanged so that waves at a single frequency in both $\mathbf{F}$ and in $\mathbf{F_c}$ can appear to travel together in phase. However $\mathbf{F_c}$ are now left-handed fields propagating on the 'advanced' light cone representing energy that has to be generated in the future in order to match boundary conditions in the past (or present). This means that $\mathbf{F_c}$ act as 'advanced' waves that were recognised by Wheeler and Feynman [**41**]. These waves can be said to carry 'negative' energy that is linked to their negative frequency. Such waves do not obey the usual temporal laws where 'cause precedes the effect' and need special consideration. A brief outline is given below.

## 8. Photon-like plane wave-packets

The simplest wave packet with a transverse vector potential $\mathbf{A}_\perp$ (lying for example in the $Ox_1x_2$ plane) is given from two (right-handed) waves propagating together but with different frequencies:

$$\mathbf{A}_\perp = \mathbf{A_N} \exp[-i\, 2\pi f_N(t_3 - x_3)] - \mathbf{A_N} \exp[-i\, 2\pi f_g(t_3 - x_3)] \tag{8.1}$$

Here if $\mathbf{A}_\perp$ vanishes at both $(t_3 - x_3) = 0$ and $(t_3 - x_3) = \tau_0$, one finds:

$$\mathbf{A}_{\perp \tau_0} = -2i\, \mathbf{A_N} \exp[-i\pi(f_N + f_g)(t_3 - x_3)] \sin[\pi(f_N - f_g)(t_3 - x_3)] = 0 \tag{8.2}$$

yielding $(f_N - f_g)\tau_0 = N$, a Lorentz invariant integer. Energy is trapped within an interval $\Delta(t_3 - x_3) = \tau_0$ moving at the speed of light. Fields outside this interval may be zero without violating Maxwell's equations. However there are no restrictions on $\tau_0$ or $f_g$ and $f_N$. This type of wave-packet is a broken crutch if it is leant on to explain quantised wave-packets.

Left-handed waves propagating together with right-handed waves dramatically change this picture. Assume that the right-handed bi-vectors are described in terms of the vector potential components $\mathbf{A}^\wedge_\perp$ propagating in the unit direction $\mathbf{k}^\wedge$ (parallel with $Ox_3$) with frequency $f_N$. The 'right-handed' bivector plane-wave fields are then:

$$\boldsymbol{\gamma_s} \cdot \boldsymbol{\gamma_t}[\mathbf{F}^\wedge(\mathbf{k_N})] = \boldsymbol{\gamma_s} \cdot \boldsymbol{\gamma_t}[ik_N\{\mathbf{A}^\wedge_\perp + I[\mathbf{k}^\wedge \times \mathbf{A}^\wedge_\perp]\} \exp[-i\, 2\pi f_N(t_3 - x_3)]] \qquad (k_N = 2\pi f_N)$$

$$(8.3)$$

The 'left-handed' bi-vector plane-wave fields with a distinct frequency $f_g$ require the signs of $i$ and $I$ to be changed:

$$\boldsymbol{\gamma_s} \cdot \boldsymbol{\gamma_t}[\mathbf{F}^\wedge_c(\mathbf{k_g})] = \boldsymbol{\gamma_s} \cdot \boldsymbol{\gamma_t}[-ik_g\{\mathbf{A}^\wedge_{c\perp} - I[\mathbf{k}^\wedge \times \mathbf{A}^\wedge_{c\perp}]\} \exp[i\, 2\pi f_g(t_3 - x_3)]] \qquad (k_g = 2\pi f_g)$$

$$(8.4)$$

Here $k_N \mathbf{A}^\wedge$ and $k_g \mathbf{A}^\wedge_{c\perp}$ determine the $\mathbf{E}$-fields while $I\, k_N \mathbf{k}^\wedge \times \mathbf{A}^\wedge_\perp$ and $I\, k_g\, \mathbf{k}^\wedge \times \mathbf{A}^\wedge_{c\perp}$ determine the $\mathbf{B}$-fields. Note that as in equation (8.1) both waves in equations (8.3) and (8.4) move together with the same phase velocity and consequently can form a wave packet confining the electromagnetic fields similar to the waves in equation (8.1). But now the different handedness of the waves means that the $\mathbf{B}$-fields can be zero while the $\mathbf{E}$-fields are a maximum at the edges of the packet. This is seen as follows. Consider the net bivector $\boldsymbol{\gamma_s} \cdot \boldsymbol{\gamma_t}[\mathbf{F}^\wedge(\mathbf{k_N}) + \mathbf{F}^\wedge_c(\mathbf{k_g})]$ with the net $\mathbf{B}$-fields, given by the coefficient of $I$, set to zero at ($t_3 - x_3$) = 0 and also at ($t_3 - x_3$) = $\tau_0$ with appropriately matching the polarisations and amplitudes so that $k_N \mathbf{A}^\wedge_\perp = -k_g \mathbf{A}^\wedge_{c\perp} = \mathbf{A}_1$ and

$$0 = \mathbf{k}^\wedge \times \mathbf{A}_1 \exp(-i\, 2\pi f_N\, \tau) - \mathbf{k}^\wedge \times \mathbf{A}_1 \exp(i\, 2\pi f_g \tau) \quad , \quad [\tau = \tau_0, 0] \qquad (8.5)$$

This equation (8.5) shows that the **B**-field is zero at the edges of the wave-packet [($t_3 - x_3$) = 0, $\tau_0$] and consequently, unlike the Fabry-Perot resonator [**51**], no currents are required in any mirrors to confine the magnetic fields to the interior of the wave-packet. With the resonant **B**-fields zero, the E-fields have to be a maximum at the edges of such a resonant wave packet so that:

$$|\exp(-i\, 2\pi f_N\, \tau_0) + \exp(i\, 2\pi f_g \tau_0)| = 2 \qquad (8.6)$$

Confining the fields to the interval $0 \leq (t_3 - x_3) \leq \tau_0$ means that **E** suddenly falls to zero outside the interval. However this does not violate the Maxwell boundary condition requiring that $\partial_s \cdot \mathbf{E} = 0$, because for this plane wave **E** is normal to both $Os_3$ and **k**.

The left- and right-handed plane wave fields, propagating together with the same phase velocity but different frequencies, now provide a unique resonant wave-packet travelling at the speed of light. It is not difficult to show that a Lorentz transformation for (**E**, **B**), together with the conjugate Lorentz transformation for ($\mathbf{E_c}$, $-\mathbf{B_c}$), preserve the boundary conditions  In other words one can form a Lorentz invariant wave-packet from left and right-handed plane waves moving with the same phase velocity that traps energy without violating any boundary conditions.

At first sight this seems no more remarkable than the conventional wave-packet but consider a wave-packet with the smallest possible value of $f_g$. The boundary conditions require that $|\exp(i\, 2\pi f_g \tau_0)| = 1$ and hence for the minimum frequency require that the wave-packet is half a wave-length long giving:

$$2\pi f_g \tau_0 = \pi \;\; ; f_g = 1/(2\tau_0) \;\; ; \;\; \exp(i\, 2\pi f_g \tau_0) = -1 \qquad (8.7)$$

The value $f_g = 0$ is not a realisable option because this gives $\partial_{t3}\mathbf{A}_\perp = 0$ which leads to zero E-fields everywhere. Having chosen $f_g$, equation (8.7) shows that the eigen frequencies for the moving resonator require $f_N$ to satisfy:

$$\exp(-i\, 2\pi f_N \tau_0) = -1 \;\; ; \qquad (8.8)$$

$$2\pi f_N \tau_0 = (2N+1)\pi \; ; \; f_N = (N+\tfrac{1}{2}) f_O \; ; \; f_O = 2 f_g \tag{8.9}$$

These are the characteristic frequencies of a photon with a fundamental frequency $f_O$.

It is now possible to see a clear rationale for choosing the lowest possible value of $f_g$. This is the frequency of the left-handed wave on the advanced branch of the light-cone. In other words this waves carries energy from the future into the past so as to enable the present (or past) boundary conditions to be met. Making the quantum ansatz that frequency and energy are related in the usual way through Planck's constant means that $f_g$ must be kept as small as possible to ensure the greatest stability of this photon-like wave-packet with a net transferable energy of $h(f_N - f_g) = N\,h\,f_O$. Such a wave-packet does not violate causality because $N \geq 0$ so that it carries only non-negative energy from the past into the future: a thoroughly causal process.

A further paper is in preparation detailing the behaviour of this type of wave-packet both within a waveguide as well as free space and with possible applications to quantum communication [**52**].This brief outline of how quantised frequencies appear within this theory is included here for two reasons. First, Dvoeglazov has remarked that additional dimensions would be required to demonstrate quantisation [**7**]. Second, it seemed desirable to show clear directions in which this new theory could take by demonstrating how quantisation of electromagnetic fields might be caused directly from boundary conditions in spacetime.

## 9. Conclusions

A new theory using Geometric Algebra is described which derives Maxwell's equations with fields, charges and potentials entirely from the symmetry of a '3+1' spacetime embedded within a '3+3' spacetime. The '3+3' spacetime has a preferred transverse temporal plane so as to remove the major difficulties previously identified with homogeneous three-dimensional time [**21,22**]. Charges appear as poles and zeros within the transverse temporal co-ordinates viewed as a complex plane. Averaging over transverse time with a contour integral removes the observational dependence on transverse time and leads to real Maxwellian fields and charges. The plane wave fields have **E**, **B** and the wave-vector **k** forming a right-handed set of vectors propagating along the 'retarded' light cone exactly as usual. However a conjugate set of fields can be formed with exactly the same overall chirality of spacetime but with $\mathbf{E_c}$, $\mathbf{B_c}$ and wave-vector $\mathbf{k}$ forming a left-handed set of vectors, moving with the same phase velocity as the conventional set now. The frequency and wave-vector change sign. These waves appear to have 'negative' energy, or energy travelling 'backwards' in time propagating along the 'advanced' light cone. Lorentz invariant wave-packets can be formed by these two types of wave, trapping electromagnetic energy moving forward at the speed of light and in such a way that causality is not violated. These resonant wave-packets have the characteristic photon frequencies. The theory adds support the view of Cramer [**9**] that the quantum theory of photons has signals travelling forward and backward in time. The component of the wave–packet travelling backwards in time may be significant in helping to understand 'entanglement'.

There is much to be done to extend the theory. The resonant combinations of right and left handed plane wave-packets need extending to confined wave-packets and wave-packets within media such as an optical fibre. Another extension is to formulate the Dirac equation using similar methods and so demonstrate the electron/electromagnetic field interactions.

**10. Acknowledgements**

The author is indebted to C.Doran and A.Lasenby for access to lecture notes on GA, now in book form (Doran and Lasenby 2003). The author is indebted to E.A.B.Cole and J.A.Baldwin for discussions of early drafts.

**Appendix A. Maxwell's equations in '3+1' STA**

Time-space unit 1-vectors are $\boldsymbol{\gamma} = \{\gamma_t, \boldsymbol{\gamma_s}\}$ where $\boldsymbol{\gamma_s} = \{\gamma_{s1}, \gamma_{s2}, \gamma_{s3}\}$. The co-ordinate 1-vector (C. = 1 units) is

$$\underline{\mathbf{X}} = \mathbf{X}.\boldsymbol{\gamma} = \gamma_t\, t + \boldsymbol{\gamma_s}.\mathbf{x} = \gamma_t\, t + \Sigma_n\, \gamma_{sn}.x_n \tag{A.1}$$

with a time-like metric: $\underline{\mathbf{X}}\,\underline{\mathbf{X}} = t^2 - \mathbf{x}.\mathbf{x}$. The commutation rules for the unit orthogonal 1-vectors require:

$$\gamma_{sm}\gamma_{sn} = -\gamma_{sn}\gamma_{sm}\ (m \neq n)\,;\ \gamma_t\gamma_{sn} = -\gamma_{sn}\gamma_t,\ \gamma_t^2 = 1,\ \gamma_{sm}^2 = -1,\ (\text{A.ll m,n}); \tag{A.2}$$

Given $\underline{\mathbf{V}}$ and $\underline{\mathbf{W}}$ as two arbitrary 1-vectors $\underline{\mathbf{V}} = \boldsymbol{\gamma_s}.\mathbf{v_s} + \gamma_t v_t$ and $\underline{\mathbf{W}} = \boldsymbol{\gamma_s}.\mathbf{w_s} + \gamma_t w_t$. Then using only equations (A.2):

$$\underline{V}\,\underline{W} = -\mathbf{v_s}\cdot\mathbf{w_s} + v_t w_t - \varphi_s\,\boldsymbol{\gamma_s}\cdot(\mathbf{v_s}\times\mathbf{w_s}) + \gamma_t\boldsymbol{\gamma_s}\cdot(v_t\mathbf{w_s} - w_t\mathbf{v_s}) \tag{A.3}$$

where $\varphi_s = (\gamma_{s1}\gamma_{s2}\gamma_{s3})$ with $\varphi_s^2 = 1$ and $\varphi_s\gamma_{s1} = \gamma_{s2}\gamma_{s3}$ etc. The term $\varphi_s\,\boldsymbol{\gamma_s}\cdot(\mathbf{v_s}\times\mathbf{w_s})$ is a purely spatial bi-vector, while $\gamma_t\boldsymbol{\gamma_s}\cdot(v_t\mathbf{w_s} - w_t\mathbf{v_s})$ is is a spacetime bi-vector. Further geometric products gives four types of 3-vector (or 'pseudo-1-vector') spanned by $\{I\,\gamma_t\,;\,I\,\boldsymbol{\gamma_s}\}$ with $I$ as the 'pseudo-scalar' defined from the ordered product:

$$I = \gamma_t\,\gamma_{s1}\gamma_{s2}\gamma_{s3}\,; \qquad I^2 = -1 \tag{A.4}$$

The spacetime 1-vector differential is denoted here by

$$\Box = (\partial_t\gamma_t - \boldsymbol{\partial_s}\cdot\boldsymbol{\gamma_s}) \tag{A.5}$$

Equations (A.3) and (A.6) enable a wave equation $(\partial_t^2 - \partial_{s1}^2 - \partial_{s2}^2 - \partial_{s3}^2)\Phi = 0$ to be written as:

$$\Box\,\Box\,\Phi = 0 \tag{A.6}$$

Defining a 1-vector $\underline{A} = (\gamma_t\,a_t + \boldsymbol{\gamma_s}\cdot\mathbf{a_s})$ then the methods of equation (A.3) evaluates $\underline{\Delta}\,\underline{A}$:

$$\Box\,\underline{A} = (\partial_t\gamma_t - \boldsymbol{\partial_s}\cdot\boldsymbol{\gamma_s})(\gamma_t\,a_t + \boldsymbol{\gamma_s}\cdot\mathbf{a_s}) = \Phi + \underline{F}. \tag{A.7}$$

where $\Phi = [\partial_t\,a_t + \boldsymbol{\partial_s}\cdot\mathbf{a_s}]$ is a scalar and $\underline{F} = \boldsymbol{\gamma_s}\,\gamma_t\cdot\mathbf{F}$ is a bi-vector with

$$\mathbf{F} = [-\boldsymbol{\partial_s}\,a_t - \partial_t\,\mathbf{a_s} + I(\boldsymbol{\partial_s}\times\mathbf{a_s})] = \mathbf{E} + I\,\mathbf{B} \tag{A.8}$$

$$\mathbf{E} = -\boldsymbol{\partial_s}\,a_t - \partial_t\,\mathbf{a_s},\quad \mathbf{B} = (\boldsymbol{\partial_s}\times\mathbf{a_s}). \tag{A.9}$$

Traditionally the Maxwell fields are said to form a polar vector $\mathbf{E}$ and axial vector $\mathbf{B}$ [43]. Now consider the multi-vector equation:

$$\Box\,\underline{F} = \underline{J} \tag{A.10}$$

where $\underline{J} = \rho\,\gamma_t + \boldsymbol{\gamma_s}\cdot\mathbf{J}$. Again using only the commutation rules (A.3) gives

$$\boldsymbol{\gamma_s}\cdot[-\partial_t\,\mathbf{F} - I(\boldsymbol{\partial_s}\times\mathbf{F})] + \gamma_t\,\boldsymbol{\partial_s}\cdot\mathbf{F} = \rho\,\gamma_t + \boldsymbol{\gamma_s}\cdot\mathbf{J} \tag{A.11}$$

Equating the different types of multi-vector on each side of this equation yields

$$\mathbf{J} + \partial_t\,\mathbf{E} = \operatorname{curl}\mathbf{B}\,;\; \partial_t\,\mathbf{B} = -\operatorname{curl}\mathbf{E}\,;\; \boldsymbol{\partial_s}\cdot\mathbf{E} = \rho\,;\; \boldsymbol{\partial_s}\cdot\mathbf{B} = 0\,; \tag{A.12}$$

Equation (A.10) gives Maxwell's equations in terms of a single multi-vector equation.

**Appendix B. Spatial rotations**

Consider an example of a multivector in 3+3 spacetime:

$$\underline{\mathbf{C}} = \gamma_{s1} P_{s1} + \gamma_{s2} P_{s2} + \gamma_{s3} P_{s3} + \gamma_{t3} U + \gamma_{t1} V_T \tag{B.1}$$

Consider $\mathbf{R}_{12} = \exp(\frac{1}{2}\gamma_{s1}\gamma_{s2}\,\theta) = \cos \frac{1}{2}\theta + \gamma_{s1}\gamma_{s2} \sin \frac{1}{2}\theta$ with $\mathbf{R}_{12}^{-1} = \exp(-\frac{1}{2}\gamma_{s1}\gamma_{s2}\,\theta)$.

$\mathbf{R}_{12}\,\underline{\mathbf{C}}\,\mathbf{R}_{12}^{-1}$

$= \exp(\frac{1}{2}\gamma_{s1}\gamma_{s2}\,\theta)\,[\gamma_{s1} P_{s1} + \gamma_{s2} P_{s2} + \gamma_{s3} P_{s3} + \gamma_{t3} U + \gamma_{t1} V_T]\,\exp(-\frac{1}{2}\gamma_{s1}\gamma_{s2}\,\theta)$

$= \gamma_{s3} P_{s3} + \gamma_{t3} U + \gamma_{t1} V_T + \exp(\gamma_{s1}\gamma_{s2}\,\theta)\,[\gamma_{s1} P_{s1} + \gamma_{s2} P_{s2}]$

$= \gamma_{s3} P_{s3} + \gamma_{t3} U + \gamma_{t1} V_T + [\cos \theta + \gamma_{s1}\gamma_{s2} \sin \theta][\gamma_{s1} P_{s1} + \gamma_{s2} P_{s2}]$

$= \gamma_{s3} P_{s3} + \gamma_{t3} U + \gamma_{t1} V_T + \{\gamma_{s1}\,(C.\!\cos \theta - P_{s2} \sin \theta) + \gamma_{s2}\,(P_{s2} \cos \theta + P_{s1} \sin \theta)\}$ (B.2)

The rotor pair $\{\mathbf{R}_{12}, \mathbf{R}_{12}^{-1}\}$ rotates multi-vectors that lie in the plane $\gamma_{s1}\gamma_{s2}$ by an angle $\theta$.

The 'half-angle' rotor such as $\mathbf{R}_{\mu\eta} = \exp(\frac{1}{2}\,\beta\,\gamma_\mu\gamma_\eta)$ is a feature of GA/STA. It interchanges or transforms 1-vectors that lie in the $\gamma_\mu\gamma_\eta$ plane and has no affect on 1-vectors that anti-commute with each $\gamma_\mu$ of $\gamma_\eta$. Any number of products of one vectors, such as (A. **B**… **C**), are rotated as $\mathbf{R}_{\mu\eta}$ (A. **B**… **C**) $\mathbf{R}_{\mu\eta}^{-1}$.

**Appendix C. Lorentz transformations**

Rotor pairs such as $\{\mathbf{R}_L = \exp(\frac{1}{2}\alpha\gamma_{s3}\gamma_{t3}),\ \mathbf{R}_L^{-1} = \exp(-\frac{1}{2}\alpha\,\gamma_{s3}\gamma_{t3})\}$ 'rotate' in the plane $\gamma_{s3}$ and $\gamma_{t3}$ and this actually changes multi-vectors into a moving frame with a boost parameter $\alpha$ along the $Os_3$. Such Lorentz rotors do not effect 1-vectors not lying in the plane $Os_3\,Ot_3$ and hence any multi-vectors formed from $\gamma_{t1}, \gamma_{t2}, \gamma_{s1}, \gamma_{s2}$ are unaltered. Although $\{\mathbf{R}_L, \mathbf{R}_L^{-1}\}$ can apply in '3+3' spacetime they are essentially transformations of moving frames of reference in '3+1' space time. To conserve space, the reader is referred to Doran and Lasenby for example to see how this rotor works most generally in '3+1' spacetime [**40**]. A limited example is given here with Lorentz rotors:

$$\mathbf{R_{Lk}} = \exp \tfrac{1}{2}\alpha\,\boldsymbol{\gamma_s.k}\hat{}\,\gamma_t\,;\ \ \mathbf{R_{Lk}}^{-1} = \exp -\tfrac{1}{2}\alpha\,\boldsymbol{\gamma_s.\,k}\hat{}\ \gamma_t \tag{C.1}$$

where $\boldsymbol{\gamma_s.\,k}\hat{}$ is the unit 1-vector in the direction of propagation of an e.m. wave with bivectors fields $\boldsymbol{\gamma_s.}\gamma_t\,\mathbf{F_k} = \boldsymbol{\gamma_s.}\gamma_t\,(\mathbf{E} + I\,\mathbf{B})$. The bivector $\boldsymbol{\gamma_s.\,k}\hat{}\ \gamma_t$ anti-commutes with $\boldsymbol{\gamma_s.}\gamma_t\,\mathbf{F_k}$ because $\boldsymbol{\gamma_s.F_k}$ and $\boldsymbol{\gamma_s.\,k}\hat{}$ are orthogonal and consequently $\gamma_s \cdot \gamma_t \, F_{Lk} = R_{Lk} \, \gamma_s \cdot \gamma_t \, F \, R_{Lk}^{-1} = (\exp \alpha \, \gamma_s \cdot k\hat{}\gamma_t)[\, \gamma_s \cdot \gamma_t \, (E + I\,B)]$

$= [\cosh \alpha + \gamma_s \cdot k\hat{}\gamma_t \sinh \alpha][\, \gamma_s \cdot \gamma_t \, (E + I\,B)]$

$= [\cosh \alpha \, \gamma_s \cdot \gamma_t \, (E + I\,B)] + \varphi_s \, \gamma_s \cdot k\hat{} \times (E + I\,B)]$

$= [\cosh \alpha \, \gamma_s \cdot \gamma_t \, (E + I\,B)] + \gamma_s \cdot \gamma_t \, k\hat{} \times (I\,E - B)]$ (C.2)

Finally re-arranging, the Lorentz transformed bivectors become:

$\gamma_s \cdot \gamma_t \, (E_L + I\,B_L) = \gamma_s \cdot \gamma_t \, [(\cosh \alpha \, E - \sinh \alpha \, k\hat{} \times B) + I\,(\cosh \alpha \, B + \sinh \alpha \, k\hat{} \times E)]$

(C.3)

**Appendix D. Rotations in the transverse temporal plane**

Just a $\exp(½\gamma_{s1}\gamma_{s2}\,\theta)$ determines a rotation in the $\gamma_{s1}\gamma_{s2}$ plane so $\exp(-½\gamma_{t1}\gamma_{t2}\,\theta) = \exp(-i\,½\theta)$ determines a rotation in the $\gamma_{t1}\gamma_{t2}$ plane. Using the commutation rules

$\exp(-i\,½\theta)[\gamma_{t1}\,a_{t1} + \gamma_{t2}\,a_{t2} + \gamma_{t3}\,a_{t3}]\exp(i\,½\theta)$

$= \exp(-i\,\theta)[\gamma_{t1}\,a_{t1} + \gamma_{t2}\,a_{t2}] + \gamma_{t3}\,a_{t3}$

$= \gamma_{t1}\,[a_{t1}\cos\theta - a_{t2}\sin\theta] + \gamma_{t2}\,[a_{t2}\cos\theta + a_{t1}\sin\theta] + \gamma_{t3}\,a_{t3}$ (D.1)

Again because $\gamma_{t3}$ lies outside the $\gamma_{t1}\gamma_{t2}$ plane, the term $\gamma_{t3}\,a_{t3}$ is unaltered. The variations in principal time given from $\exp(-i\,kt_3)$ and $\exp(+ikt_3)$ provide rotations in the transverse temporal plane and also provide an arrow of time by keeping $k > 0$.